# Rod-ring paradox

Øyvind Grøn[¤] and Eirik Berntsen[#]

¤ Øyvind Grøn, Oslo Met, P. O. Box 4, St Olavs Plass, 0130 Oslo, Norway

# Eirik Berntsen, Enebakkveien 25a, 0657 Oslo, Norway

**Abstract**

A new apparent relativistic paradox is presented involving only one space-time event. This is different from earlier 'relativistic paradoxes' involving extended bodies or events at different positions. A collision between a rod and a ring impacting at an oblique angle to each other is considered in the context of the special theory relativity. A question arises as to where along the length of the rod the point of impact will be according to two observers following the rod and the ring, respectively. The observers argue from a purely kinematical point of view in favor of two different points of impact along the rod. However there can only exist one point of impact. In order to solve this apparent paradox we use the asynchronous formulation of relativistic kinematics in which the consequences of the relativity of simultaneity are built into the formalism. We show that this reconciles the descriptions from the two frames of reference, and hence that the new 'paradox' leads to a strong argument for the relevance of the asynchronous formulation of relativistic kinematics.

1. **Introduction**

During the years since the special theory of relativity was presented [1] in 1905 several special relativistic paradoxes have been discussed. We will give a short review of those related to the one presented in this paper.

• *The twin paradox*. This is the most notorious of the special relativistic paradoxes. There have been published several hundred articles about it. Some recent references are [2-13]. The paradox is that if two twins travel away from each other and later come back to each other again, then if each twin considers himself as at rest, they will both predict that their brother is younger than themselves when they meet again. The paradox arises because only the special relativistic time dilation is used by the twins to predict the ageing of their brother during the time they are separated, while its formulation requires the general principle of relativity valid for accelerated motion, since at least one of the twins must be accelerated in order that they can depart from each other and later meat again. Its solution requires that one also takes the gravitational time dilation into account [9].

• *The lever paradox*. In 1909 G. N. Lewis and R. C. Tolman [14] considered a right angled lever acted upon by forces at a right angle to each other. The forces are adjusted so that the torques are of equal magnitude, and the lever is in equilibrium in it rest frame. A Lorentz transformation to another inertial frame gives the result that the torques are unequal. There is a non-vanishing external torque, so the lever should start rotating.

This paradox has been solved in several ways. One is to introduce an energy current in the lever in the frame where the lever moves and the torques are different. This energy current provides an internal torque which cancels the external torque and secures rotational equilibrium. This solution of the paradox was introduced by Max von Laue [15] already in 1911, and the energy current in the lever arms is therefore called the *Laue current*. The second solution is to give a covariant definition of torque [16]. This effectively takes account of the relativity of simultaneity in this situation. A third way is to take into account the torque due to internal forces in the lever arms without invoking the Laue current. Nickerson and Mc Adory [17] demonstrated that this internal torque cancels the external one, securing equilibrium in an arbitrary frame if there is equilibrium in the rest frame of the lever. Finally the paradox has been solved by introducing an *asynchronous description* of the lever in the frame where it moves [18-20]. J. Güémez has recently given a comprehensive analysis [21] of the lever paradox using the asynchronous formulation of relativistic statics.

At this point a few words about this formalism are in order. Just as using differential forms instead of ordinary tensors gives a stronger formalism by building antisymmetry into the formalism, the asynchronous formalism is stronger than the usual synchronous formalism because effects of the relativity of simultaneity are built into the formalism.

The main point of the asynchronous formalism is to define a physical body as a set of simultaneous events in the rest frame of the body. Hence the body consists of a set of non-simultaneous events in a frame where it moves. That is the reason for the term 'asynchronous'.

None of the apparent paradoxes mentioned in this introduction appear when the asynchronous formalism is used.

• *Length contraction paradoxes.* There are several versions of this paradox. We shall consider them under two headings.

LCP1. *The falling rod paradox*. This was introduced by W. Rindler [22] in 1961 and has later been analyzed by several researchers [23-28]. The seeming paradox is solved by taking into account the relativity of simultaneity, and will be analyzed in section 2.

LCP2. *The ladder paradox*. Another version of the length contraction paradox is called the ladder paradox or the pole in the barn paradox [29]. In one version a ladder moves into a barn (or a tunnel) and due the length contraction there is place for it in the barn, so the doors at the two ends of it can be closed with the rod inside the barn. However, as described from the rest frame of ladder the barn is Lorentz contracted and the ladder not, so the barn is too short to contain the ladder. A related problem involving a lock and a key has been considered by E. Pierce [30]. These paradoxes are solved by taking the relativity of simultaneity into account.

The paradox presented in the abstract is different from the previous ones. It is the first special relativistic paradox which in a given situation involves only one point event: the event that the rod and the ring would have first touched each other if they had moved in the same plane. Such spacetime events are relativistic invariants. Hence if the event produces a collision mark in the rod, there cannot be two marks at different positions in two different inertial frames. Yet this is the result of the traditional analysis describing the rod and the circle by simultaneity in the two frames. Thus a seeming paradox appears.

The analysis provides a new example showing the significance of the relativity of simultaneity in the special theory of relativity. It demonstrates that the usual description of two bodies moving relative to each other, by simultaneity in the rest frames of the two bodies, relate two different sets of events, and in this sense represent two different physical situations.

## 2. Principle for deciding what happens to an extended body and a reanalysis of the Rindler length paradox

We shall need a definition of the concept 'physical body' for bodies moving in a Born rigid way so that the rest length between their different parts does not change.

**Definition 1**: A *physical body* is a set of events that are simultaneous in the rest frame of the body.

This definition is a consequence of defining space as a set of simultaneous events. Hence the spatial extension of a physical body is found by considering a set of events that are simultaneous in the rest frame of the body.

Due to the relativity of simultaneity this definition implies that a moving physical body consists of events that are not simultaneous. Hence it implies an asynchronous description of physical bodies [18-20].

Also we shall need a rule in order to know what happens to a physical body under given circumstances. Traditionally the following rule has been used.

**Rule 1**. *To know what happens to a physical body in a physical situation we must describe the participating bodies as sets of events that are simultaneous in the laboratory frame*.

This definition has some strange consequences. Let us consider the Rindler length contraction paradox [22]. Rindler considered the following situation: A rod sliding on a horizontal table comes to a hole with the same rest length as the rod. In the laboratory, where the rod moves, it is Lorentz contracted so that it is shorter that the hole, and it falls down. But in the rest frame of the rod the hole is Lorentz contracted, and the rod is longer than the hole, so it seems that it cannot fall down.

Rindler solved the paradox by implicitly using the rule 1 and stated that "there can be no doubt that the description in the laboratory is correct". Thus Rindler considers the laboratory frame a preferred frame for deciding what happens to the rod.

Hence Rindled decides that the rod will fall down. And due to the relativity of simultaneity the rod will deform as described in its own rest frame. In order to illustrate what was happening Rindler provided the following figures.

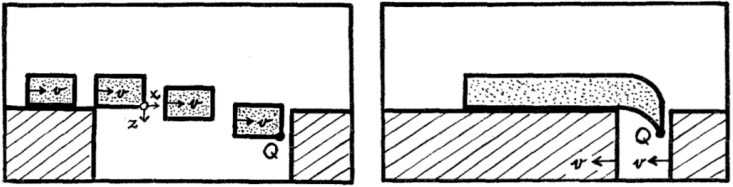

*Fig.1. Left: The situation described at four successive points of time in the laboratory frame. Right: The situation as described in the rest frame of the rod at one particular instant.*

The figure is drawn for the case that the rod has the same rest length, $L_0$, as the length of the hole, and it moves with a velocity $v$ such that $\gamma = \left(1 - v^2/c^2\right)^{-1/2} = 4$. Then the Lorentz contracted length of the rod is $L = L_0/4$, and it falls into the hole. The right hand figure is obtained by making a Lorentz transformation into the rest frame of the hole, and describing the hole by a set of events that are simultaneous in the rest frame of the hole. In this frame the hole moves to the left. The Lorentz transformation implies that simultaneous events at different points on the rod in the rest frame of the rod are transformed to a set of non-simultaneous events in the rest frame of the hole. The events to the right happen later than the events to the left. Hence the events to the right happen at lower positions as shown on the figure to the left. These lower positions are drawn at the right part of the rod in the right hand figure.

Two problems arise. 1. The answer to the question: 'does the rod bend?' depends upon the motion of the observers who gives the answer. 2. If one does not consider the laboratory frame as a preferred frame, but accepts the usual synchronous description of the situation by an observer at rest relative to the rod as equally valid, then what happens to the rod – whether it falls down in the hole or not – depends upon the motion of the observer.

*We need a rule so that what happens to a rod is Lorentz invariant*, i.e. it is independent of the velocity of the observer. This rule takes the form.

**Rule 2**. *In order to know what happens to a physical body in a physical situation we must describe the participating bodies as sets of events that are simultaneous in the rest frame of the body*.

Such a rule was hinted at by W. H. Wells [31] in a comment to the length contraction paradox where he writes: "Since the elastic theory of solids is nonrelativistic, questions as to the shape of a body must be decided in the inertial frame where the body is instantaneously at rest."

Note that the rule 2 in no way implies that the rest frame of the body is preferred for the description of what happens. But it will often be the simplest frame to use for deciding what happens to a body in a given situation, since only in this frame is the body represented by a set of simultaneous events. Below we shall show that both an observer at the rest in the frame of the rod and another observer at rest in the frame of the ring, reach the same result when they use the rule 2 for deciding where the collision mark on a rod will appear. This method secures covariance in the relativistic description of extended bodies.

Also we need a definition of what we mean by the term 'physical situation'.

**Definition 2:** A *physical situation* is a given set of events.

Let us now reconsider Rindler's length contraction paradox. *We have two physical situations*. The laboratory observer at rest with the hole considers a moving rod described by events that are simultaneous in the laboratory frame. This is *situation 1*. As described in the laboratory frame the rod is Lorentz contracted and the hole not, and the rod falls down. This is the situation described by Rindler. Describing this situation in the rest frame of the rod, he gives an asynchronous description of the rod, i.e. describes it by a set of non-simultaneous events. According to this description the rod is bent and able to fall down in the hole, in spite of the fact that in this frame it is longer than the Lorentz contracted hole. This is how the observer in the rest frame of the rod explains that the rod falls down in the hole in this situation.

The paradox arises when confronting this with the description as a set of simultaneous events as given by an observer following the rod. This is *situation 2*. In this situation the moving hole is Lorentz contracted, the rod remains straight and does not fall into the hole. The description of this situation from the rest frame of the hole is as follows. A Lorentz transformation of the simultaneous events in the rod frame that define the physical rod, leads to a non-simultaneous description of the physical rod in the rest frame of the hole. The events at the front end of the rod happen later than those at the hind end. Hence the physical rod is *Lorentz lengthened* in the rest frame of the hole. This is the explanation of the laboratory observer of why the rod does not fall into the hole in this situation.

H. van Lintel and C. Gruber [28] also arrived at the conclusion that the rod will not fall through the hole arguing from a consideration of stress propagation in the rod-material.

In 1962 a new version of the paradox was presented by R. Shaw [23]. Here a horizontal plane with a hole moves vertically, and a horizontal rod moves horizontally in the laboratory plane, as shown in Figure 2 which is from reference 23.

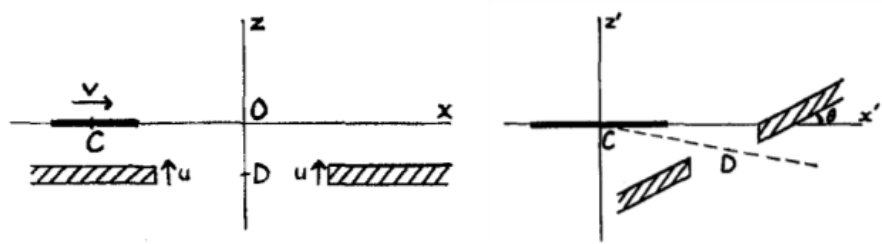

**Figure 2.** *Left: A rod moves horizontally and a plane with a hole vertically in the laboratory frame. Right: The plan makes an angle with the horizontal direction in the rest frame of the rod, and moves along the punctuated line in the figure.*

Here a horizontal plane with a hole moves vertically, and a horizontal rod moves horizontally in the laboratory plane, as shown in Figure 2 which is from reference 23. In this version the rod passes through the hole even if it is longer that the hole in its own rest frame. But due to the relativity of simultaneity the physical plane is not horizontal in the rest frame of the rod, but makes an angle with horizontal direction making it possible for the rod to pass through the hole.

A related situation in which a horizontally oriented rod has both a vertical and horizontal motion while the table with a hole is at rest, has been analyzed by C. Iyer and G. M. Prabhu [32].

Among the length contraction paradoxes that have been discussed earlier, the one presented by C. Iyer and G. M. Prabhu in ref. 33 is most similar to the one discussed in the present paper. But there is an important difference. They considered a collision of two parallel rods, D and B, inclined by the same angle relative to the horizontal direction in their respective rest frames M and K. The situation is shown in Figure 3.

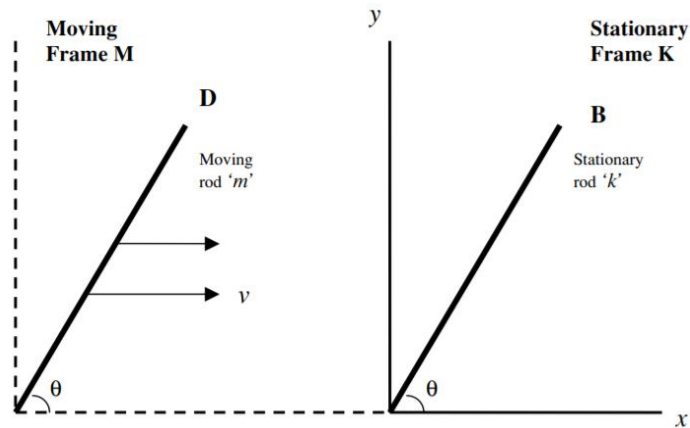

**Figure 3**. *The situation considered by C. Iyer and G. M. Prabhu.*

As described in the rest frame of D the rod B is steeper than D and therefore the collision happens first at the upper points of the rods. But as described in the rest frame of B the rod D is steeper than B, and hence the collision happens first at the lower end of the rods. These conclusions seem contradictory. What happens? Does the collision happen first at the upper or lower end of the rods?

The solution is that the concept 'first' is not Lorentz invariant for events happening at different positions due to the relativity of simultaneity. The paradox is concerned with the fact that two events at different positions in the direction of motion of a moving object that are simultaneous in a certain reference frame, may have different successions in two different inertial frames.

Iyer and Prabhu noted that the rods are observed to be parallel in a frame F which moves with the velocity $u = v/(1+1/\gamma)$ with respect to K and $-u$ with respect to M. In this frame the rods are observed to move with the same velocity $u$ against each other, and collide in such a way that all the points of the rod hit each other simultaneously. Consider now the simultaneous collisions at the bottom of the rods and the top. These events, $P_1$ and $P_2$, happen at different positions in the direction of motion of the rods. $P_1$ happens first in the frame K, and $P_2$ happens first in the frame M.

As mentioned above, we shall now consider a related, but different paradox which in a given situation involves just one event in spacetime. The question is: Where does the shadow of an inclined rod moving towards a ring first coincide with a point on the shadow of the ring? As described in the rest frames of the rod and the ring this coincidence seems to happen at two different places. The impossibility of this can be seen by imagining that a mechanism makes a mark on the rod at the coincidence of the rod and the ring, corresponding to the initial point of contact if they had moved in the same plane. Since this coincidence is a single spacetime event which is Lorentz invariant, there cannot be two marks on the rod at two different positions.

3. **Rod-ring coincidence**

Some time ago one of us (E.B.) began thinking about a relativistic collision between a ball and a plane. We here simplify the thought experiment to a coincidence between a rod and a ring travelling towards each other in the *x*-direction, being slightly separated from each other in the z-direction so that they do not collide. This arrangement is chosen in order to avoid problems with the dynamics of the rod and the ring during a collision which is irrelevant to the kinematical character of the apparent paradox presented here. We shall consider a purely kinematical problem not involving any deformations of the rod or the ring. Even if the rod and the ring are slightly behind each other, when considered in a direction normal to the planes they move along, we will use the word 'coincidence' when the shadow of the rod on a screen (parallel to the planes of motion) behind the rod and the ring, hits the shadow of the ring.

In Figure 4 the rod and the ring are drawn as they would appear on the screen if both were at rest. The radius $R$ of the rod and the angle $\theta_0$ are the given quantities.

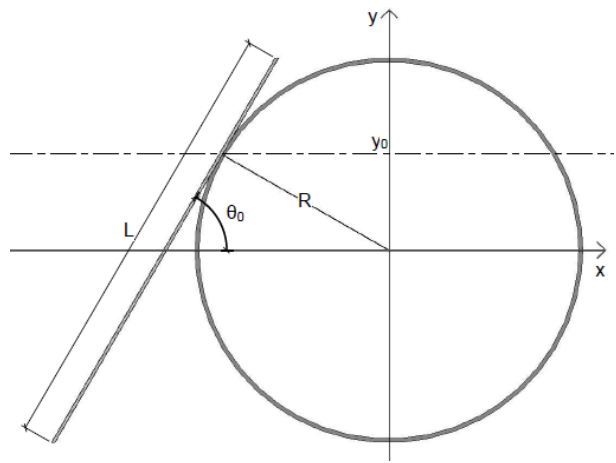

*Figure 4. The rod and the ring at rest. L is the rest length of the rod.*

We see from Figure 4 that then the ring would coincide with the rod at a height

$$y_0 = R\cos\theta_0. \tag{1}$$

In the actual situation as described from the rest frame of the rod with coordinates $(x, y)$, the ring approaches the rod with a velocity $v$. Then the ring is somewhat flattened due to the Lorentz contraction and has the shape of an ellipse,

$$\frac{x^2}{1-v^2/c^2} + y^2 = R^2. \quad (2)$$

Hence the shadow of the Lorentz contracted ring would hit the shadow of the rod at a position $y > y_0$. This is illustrated in Figure 5.

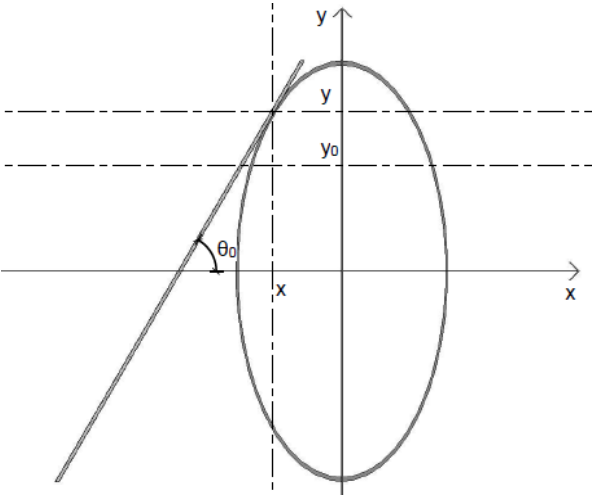

**Figure 5.** *The coincidence as seen from a frame of reference following the rod. We have here chosen $\gamma = 1/\sqrt{1-v^2/c^2} = 2$. An observer would here expect that the shadow of the ring hits the shadow of the rod at a distance y from the x-axis. Here $y_0$ is the hitting point in the limit of vanishing velocity. The slope of the rod is $\tan\theta_0$.*

As described in the rest frame of the ring with coordinates $(x', y')$, the shape of the ring is circular, and the rod is Lorentz contracted in the direction of motion as shown in Figure 6.

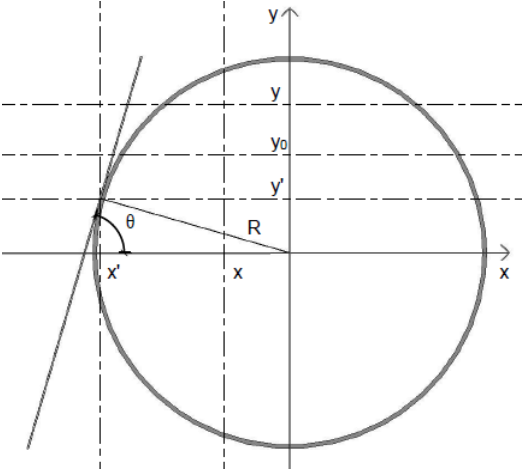

**Figure 6.** *The collision of the shadows of the rod and the ring as seen from the rest frame of the ring with γ=2. An observer following this frame of reference would expect that the shadow of the ring hits the shadow of the rod at a distance y' from the x'-axis.*

Hence the rod makes a larger angle $\theta > \theta_0$ with the $x'$-axis, and its shadow is therefore hitting the ring at a position $y' < y_0$. Therefore $y > y'$ which means that as described in the rest frames of the rod and the rest frame of the ring, the shadow of the rod is hit by the shadow of the ring at two different places. If a mechanism gives the rod a mark at this coincidence, there would appear two marks on the rod at different places according to the observers in the rest frames of the rod and the ring. Due to the Lorenz invariance of a space time event – that the shadow of the rod first touches the shadow of the ring – this cannot be the case. This is the *rod-ring paradox*.

4. **Description in the rest frame of the rod**

We shall calculate the point where the shadow of a ring with an elliptic shape due to the Lorentz contraction hits the shadow of the rod. At this point the rod and the ring have the same slope. Differentiating eq.(2) we find the slope of the Lorentz contracted ellipse,

$$\frac{dy}{dx} = -\frac{x}{(1-v^2/c^2)y}. \tag{3}$$

From eq.(2) we have at the left hand side of the ellipse

$$x = -\sqrt{1-v^2/c^2}\sqrt{R^2 - y^2}. \tag{4}$$

Hence

$$\frac{dy}{dx} = \frac{\sqrt{R^2 - y^2}}{\sqrt{1-v^2/c^2}\, y}. \tag{5}$$

Putting the slope of the ellipse equal to the slope of the rod (Figure 5) at the collision point $(x_1, y_1)$ we get

$$\frac{\sqrt{R^2/y_1^2 - 1}}{\sqrt{1-v^2/c^2}} = \tan\theta_0. \tag{6}$$

Equations (1) and (6) give

$$y_1 = \frac{R\cos\theta_0}{\sqrt{1-\frac{v^2}{c^2}\sin^2\theta_0}} = \frac{y_0}{\sqrt{1-\frac{v^2}{c^2}\sin^2\theta_0}} \tag{7}$$

showing that $y_1 > y_0$.

5. **Description in the rest frame of the ring**

In this frame the rod with rest length $L$ is Lorentz contracted in the direction of motion. Hence its slope is

$$\tan\theta' = \frac{L\sin\theta_0}{L\sqrt{1-v^2/c^2}\cos\theta_0} = \frac{\tan\theta_0}{\sqrt{1-v^2/c^2}}.$$ (8)

This shows that the rod makes a larger angle with the *x*-axis the faster it moves. Similarly to Eq.(1) the $y'-$ coordinate of the touching point is

$$y'_1 = R\cos\theta'.$$ (9)

Using equations (8), (9) and (1) this coordinate is found to be

$$y'_1 = \sqrt{\frac{1-v^2/c^2}{1-(v^2/c^2)\cos^2\theta_0}}\, y_0,$$ (10)

showing that $y_1' < y_0$. Hence $y_1 > y_1'$. Furthermore, the Lorentz transformation of the *y*-coordinate is $y' = y$. So as described from the rest frame of the rod and the rest frame of the ring the ring touches the rod at two different places. Due to the Lorentz invariance of the *y*-coordinate this is not only a coordinate effect. *The physical marks on the rod made at the coincidence are predicted by the observers in the rest frames of the rod and the ring to be at two different places on the rod.* If the two observers describe one and the same situation, this cannot be correct. This disagreement between the rod-and ring-observer is the kinematical rod-ring paradox.

**6A. Asynchronous description of the rod in the rest frame of the ring**

The first step in removing the contradiction is to note that *the physical rod is defined by sets of events that are simultaneous in its own rest frame*. Hence it is represented as a set of non-simultaneous events in the rest frame of the ring.

Thus in order to determine where the shadow of the moving rod first touches the shadow of the ring at rest, we must give an *asynchronous description* [18-20] of the rod in the rest frame of the ring. Let S denote the rest frame of the rod and S' the rest frame of the ring. Making a Lorentz transformation from S to S' we find that the set of simultaneous events in S defining the physical rod, are such that the event at the lover point of the rod happens first in S', and the events defining the physical rod happen later the closer to the front end of the rod they are. Due to the motion of the rod this means that the 'asynchronous rod', consisting of simultaneous events in the rest frame of the rod, makes a smaller angle with the $x'-$ axis the faster it moves. This causes the touching point to be displaced upwards.

The Lorentz transformation from S to S' is

$$x' = \frac{x+vt}{\sqrt{1-v^2/c^2}} \;,\quad y' = y \;,\quad t' = \frac{t+(v/c^2)x}{\sqrt{1-v^2/c^2}}.$$ (11)

The physical rod is defined by simultaneous events in S, $\Delta t = 0$. Hence, from the last of Eq.(11), the event defining the front end of the rod happens

$$\Delta x' = \frac{L\cos\theta_0}{\sqrt{1-v^2/c^2}} \tag{12}$$

later than the corresponding event defining the rear end of the rod, and from the first of Eq.(11) the front end is positioned

$$\Delta x' = \frac{L\cos\theta_0}{\sqrt{1-v^2/c^2}} \tag{13}$$

in front of the rear end. This means that the physical rod, consisting of asynchronous events in the rest frame of the ring, is *Lorentz lengthened* instead of Lorentz contracted. Thus the slope of the physical rod in S' is

$$\tan\hat{\theta}' = \frac{L\sin\theta_0}{L\cos\theta_0/\sqrt{1-v^2/c^2}} = \sqrt{1-v^2/c^2}\,\tan\theta_0. \tag{14}$$

Hence

$$\cos\hat{\theta}' = \frac{\cos\theta_0}{\sqrt{1-\left(v^2/c^2\right)\sin^2\theta_0}}. \tag{15}$$

Thus the touching point of the shadow of the ring and the shadow of the physical rod is at a position

$$\hat{y}' = R\cos\hat{\theta}' = \frac{R\cos\theta_0}{\sqrt{1-\left(v^2/c^2\right)\sin^2\theta_0}}, \tag{16}$$

which is equal to $y_1$ as given in Eq.(7). Hence with the asynchronous description of the rod in the rest frame of the ring, the observers following the ring and the rod agree on the position of the first contact between the ring and the rod (Figure 7).

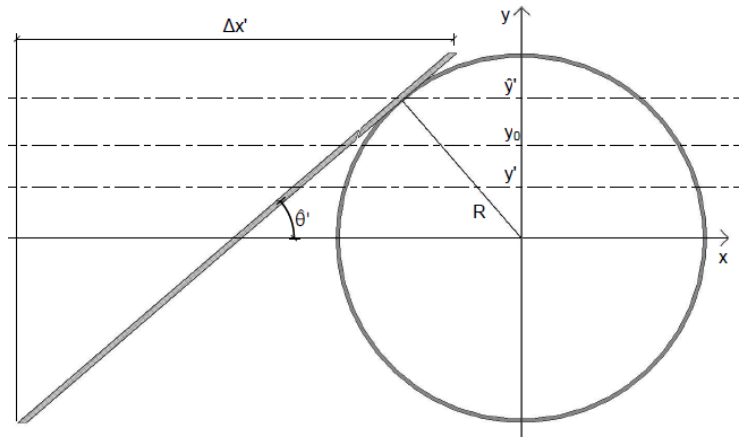

*Figure 7.* Situation with Lorentz lengthened rod. This is the physical rod as described in the ring's frame of reference, such that all points on the rod represent simultaneous events in its own frame of reference. An event A at the right most point on the rod would here occur a later time than an event B at the left end of the rod which is simultaneous to A in its own frame of reference. Notice how the rod now touches the ring at the point predicted in its own frame of reference.

### 6B. Asynchronous description of the ring in the rest frame of the rod

A second step in removing the contradiction is to note that the physical ring is defined by a set of events that are simultaneous in its own rest frame. Hence the physical ring is represented by sets of events that are not simultaneous in the rest frame of the rod. The events at the front side of the ring relative to its direction of motion happen a little time later that the events at the rear side. Hence the physical ring gets a Lorentz lengthening (Figure 8) in its direction of motion in the rest frame of the rod.

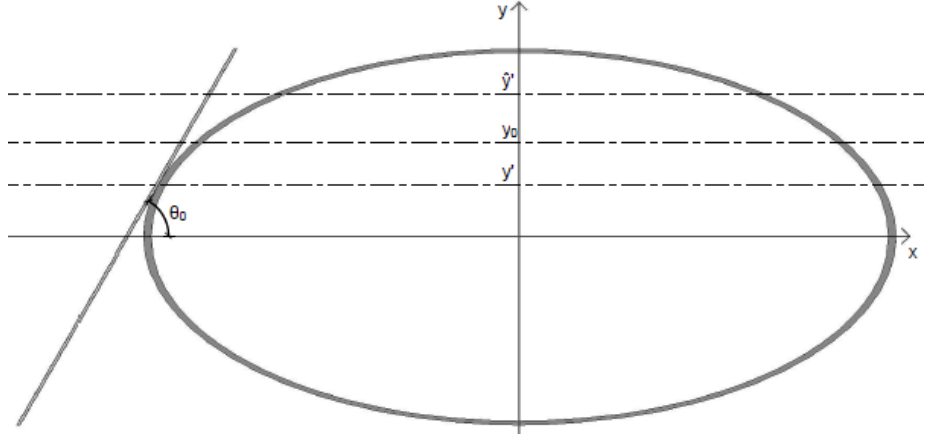

**Figure 8.** *Situation with Lorenz elongated ring, as seen from the rod's frame of reference. This is analogous to the situation shown in figure 7, but viewed from the opposite side of the collision.*

The equation of the Lorentz lengthened ring is

$$x^2\left(1-\frac{v^2}{c^2}\right)+y^2 = R^2 . \tag{17}$$

The slope of the physical ellipse is

$$\frac{dy}{dx} = -\left(1-\frac{v^2}{c^2}\right)\frac{x}{y} . \tag{18}$$

From eq.(17) we have at the front side of the ellipse

$$x = -\sqrt{\frac{R^2-y^2}{1-v^2/c^2}} . \tag{19}$$

Hence the slope is

$$\frac{dy}{dx} = \sqrt{\left(1-\frac{v^2}{c^2}\right)\left(\frac{R^2}{y^2}-1\right)} . \tag{20}$$

Putting the slope of the physical ring equal to the slope of the rod at the collision point $(x_2, y_2)$ we get

$$\sqrt{\left(1-\frac{v^2}{c^2}\right)\left(\frac{R^2}{y_2^2}-1\right)} = \tan\theta_0 . \tag{21}$$

Equations (1) and (21) give

$$y_2 = \sqrt{\frac{1-v^2/c^2}{1-\left(v^2/c^2\right)\cos^2\theta_0}}\, y_0 \tag{22}$$

in agreement with the position given in equation (10). Again the observer at rest with the rod and the one at rest with the ring agree.

We have now seen: I. An asynchronous description of the rod in the rest frame of the ring gives the same collision point as a synchronous description of the ring in the rest frame of the rod. II. An asynchronous description of the ring in the rest frame of the rod gives the same collision point as a synchronous description of the rod in the rest frame of the ring. But I and II give two different collision positions.

Now we have two consistent descriptions of the collision. But still we have two different collision positions. So where does the first touch happen?

In reality we have described two different situations. The situations I and II are not the same. The case I is a situation where the rod is considered at rest in a 'laboratory frame' where one gives a synchronous description of what happens. In this frame a Lorentz contracted ring approaches the rod and its shadow hits the shadow of the rod. The case II is a situation where the ring is considered at rest in the 'laboratory frame', and a rod, described as a successive sets of simultaneous events in the rest frame of the ring, approaches it and its shadow hits the shadow of the ring. It is not the same physical bodies that collide in the two situations.

The simultaneous set of events defining the Lorentz contracted ring in the rest frame of the rod is not the same set of events that represent the physical ring. Similarly the simultaneous set of events defining the rod with a Lorentz contraction in the direction of motion is not the same set of events that define the physical rod. Hence the physical situations defined by the synchronous descriptions in the rest frame of the ring and the rod, respectively, are not identical. That is the reason for the different positions of the collision marks with the synchronous description in the two rest frames.

7. **Conclusion**

There exist several apparent paradoxes where fast moving extended bodies are described in the context of the special theory of relativity, arising if one neglects to take the relativity of simultaneity properly into account. One is the Rindler length contraction paradox [5], also called the ladder paradox [6-11]. We have here considered a new apparent paradox of a similar type.

In one fixed situation there cannot be more than one first point of impact in a collision between the shadows of a rod and a ring as considered above. However, as described in the rest frame of the ring the collision point is at a position closer to the x-axis than as described from the rest frame of the rod. This is not only a coordinate effect. By a proper mechanism the ring may give the rod a mark at the first point of contact of the shadows of the rod and the ring. As predicted by the observers in the rest frames of the rod and the ring, using a synchronous description in each reference frame, these marks will appear at different positions on the rod, in spite of the fact that each observer would say that a mark on the rod is assumed to appear as the result of a single space time event. This is an apparent paradox.

In order to solve this apparent paradox one has to take into consideration the relativity of simultaneity. It is a relatively common situation within special relativity thought experiments to have two spatially separated events appear to be simultaneous in one frame of reference and not in another. However in the present case a similar paradox appears with only one event involved: the first point of contact of the shadows of the rod and the ring.

We have shown that the synchronous descriptions of the collision of a rod and a ring as given in the rest frames of the rod and the ring in sections 6A and 6B, respectively, represent two different physical situations.

Furthermore, we have shown that it is a consequence of the relativity of simultaneity that in order to give consistent descriptions of each situation from the rest frames of the rod and the ring, we must give an asynchronous description of the ring in the rest frame of the rod, and an asynchronous description of the rod in the rest frame of the ring.

In situation I with the ring at rest in the 'laboratory frame' we must give an asynchronous description of the rod, and in situation II with the rod at rest in the 'laboratory frame', we must give an asynchronous description of the ring *in order to secure Lorentz invariance of the descriptions*. Both situations are of course equally 'correct'. With this procedure we have shown that the collision point is higher up along the y-axis in the situation with the rod at rest in the laboratory than in the corresponding situation with the ring at rest in the laboratory.

Stated in terms of the mechanism making a mark on the rod, the solution of the paradox may be formulated as follows: A mechanism to detect the first contact could detect the touching of the objects or the shadows and could paint a mark if the slopes of the ring and the rod at this point are the same. But this mechanism must be bolted either to the rod or to the ring, conducting the measurement in the corresponding frame of reference. When fixed to the rod, the mechanism will paint its mark at event E1, when fixed to the ring, the mark will be at event E2.

Two point of a general character should be kept in mind.

   A. *In general the synchronous description of extended bodies in two different reference frames represents two different physical situations.*
   This is in agreement with the principle that the viewpoints of observers in different inertial reference frames are equally valid descriptions of the same physical situation. But what has become clear in the rod-ring collision paradox is that the situation with the rod at rest and a moving ring described as a succession of simultaneous events, is not equivalent to a situation with the ring at rest and a moving rod described as a succession of simultaneous events. In our paper it has been shown that in each situation the viewpoints of observers in the rest frames of the rod and the ring are equally valid descriptions of the same physical situation giving the same coincidence point as predicted by each observer. But the coincidence points are different for the two different situations.

> B. *A physical body represented by a set of simultaneous events in its own rest frame is Lorentz lengthened when it is observer from a reference frame in which is moves.*

If these points are kept in mind, the apparent paradoxes connected with the relativity of simultaneity will not appear. Also this way of thinking leads to a covariant description of moving, extended bodies valid in arbitrary inertial frames.

Swings P and Rosenfeld L 1937 Astrophys. J. **86** 483–6